\documentclass[12pt]{article}

\pdfoutput=1
\usepackage{hyperref,exscale,graphicx}

\begin{document}

\title{Toy observer in unitary evolution: Histories, consciousness, and state reduction} 

\author{Lutz Polley}

\date{\small Institute of Physics, Oldenburg University, 26111 Oldenburg, FRG}

\maketitle

\begin{abstract}
For a toy version of a quantum system with a conscious observer, it is demonstrated that the
many-worlds problem is solved by retreating into the conscious subspace of an entire observer 
history.  
In every step of a discretised time, the observer tries to ``see'' records of his past and 
present in a
coherent temporal sequence, by scanning through a temporal fine-graining cascade. The extreme
most likely occurs at the end of some branch, thus determining observer's world line. The
relevant neurons, each with two dimensions, are power-law distributed in number, so order
statistics implies that conscious dimension is located almost entirely in the extremal 
branch.   
\end{abstract}

\section{Introduction}

Among the various approaches to an interpretation of quantum theory, one is to regard the 
superposition principle, state vectors, and the Schr\"odinger equation as universally valid, 
and to seek a solution to the ensuing many-worlds problem \cite{Everett1973}. An old
\cite{vNeumann1932} but still relevant \cite{Donald1999,Tegmark2014} conjecture is that the 
solution should involve the physical functioning of an observer's consciousness. 
The approach taken here falls in this category.    

As to the merits of the superposition principle, 
most concepts of quantum theory, as presented in textbooks, rely on the formalism of state 
vectors. Even thermal systems, traditionally regarded as mixtures, can be 
treated as pure state vectors \cite{Tasaki1998}. 
The basic law of propagation of a particle takes an almost self-evident form
\cite{Feynman1965,Baym1978,Zee1991,Polley2001} when positions are restricted to a spatial 
lattice, and superpositions are regarded as a logical possibility. For electromagnetic fields 
to be incorporated, only the complex phases already present in the hopping amplitudes need to 
be varied \cite{Wilson1974}. In comparison, Newton's laws are phenomenological.       

A fundamental problem of a linear evolution equation emerges in application to a quantum system interacting with an observer. With a system in a superposition of properties $1,\ldots,B$, and an observer trying to determine the ``actual'' property, the Schr\"odinger equation implies a transition of the form 
\begin{equation}   \label{EqualAmplitudeSplitting}
   \left(\sum_{n=1}^B |n\rangle \right)|\mathrm{ready}\rangle \longrightarrow
   \sum_{n=1}^B \Big(|n\rangle|\mathrm{observed}\,n\rangle\Big)  
\end{equation} 
While experience suggests an observer should find himself
in one of the states $|\mathrm{observed}\,n\rangle$ after the measurement, the superposition state produced by the Schr\"odinger equation does not indicate which of the possible results has ``actually'' been obtained. The observer rather seems to have split into $B$ branches of himself. To date, no consensus exists as to whether a superposition of observer states like (\ref{EqualAmplitudeSplitting}) describes something physically real. The problem appears less dramatic when the state vector is converted into a density matrix, as in the theory of decoherence, but the ambiguity about the result of the measurement persists \cite{Schlosshauer2007}.  
In Everett's many-worlds interpretation \cite{Everett1973} the superposition \emph{is} observer's real wavefunction. His inability to realise more than one of his branches is inferred from the (undisputed) impossibility of branches interacting with each other. While certain activities like talking in branch 1 about events in branch 2 can be ruled out in this way, a mere simultaneous awareness of branches is not covered by the argument 
\cite{Penrose1997}. An attempt at resolving this problem was made by this author in 
\cite{Polley2012}; in the present paper, that approach is simplified and generalised. The hypothesis is that an observer's awareness is extremal in one branch, in such a way that the sum of remaining branches can be regarded as a negligible contribution. 

In the Copenhagen interpretation, the process of measurement is not described by 
a linear equation for amplitudes in a superposition, but is described by state reduction 
following Born's rule. Accordingly, superpositions evolve in a stochastic way. Superposition 
(\ref{EqualAmplitudeSplitting}) would end up as $|n\rangle|\mathrm{observed}\,n\rangle$ with 
probability $1/B$ for each case. The disturbing point here is that 
probabilities occur at a fundamental level---no natural law supposedly exists that would 
determine, in principle, the outcome of a single quantum measurement. 
By contrast, an agreeable role for stochastics would be that of
an approximation to deterministic but uncontrollably complicated dynamics. In the model of this
paper, there will be a constant law of evolution (representing determinism) given by a unitary
operator which is a peculiar kind of random matrix (assumed to approximate some complicated
dynamics). Pseudo-random evolution results if the operator is applied repeatedly to an
initial state of an appropriate class.

Since the early days of quantum mechanics the idea has been pondered that state reduction may involve an observer's consciousness \cite{vNeumann1932}. While the observer remains conscious with every result of the measurement, there may be variations in the degree of consciousness. Of all the details around us that could in principle catch our attention, only a tiny fraction actually does so. If entire histories are considered, that fraction multiplies with every instant of time, so there is lots of room for outstanding extremes. 

A formal description of the physical functioning of a \emph{real} observer, including both 
Everettian branching and the influence of an irreducible stochastic process in the generation
of the branches, has been given by Donald \cite{Donald1999}. For a model of state reduction,
it may be an unnecessary complication to consider consciousness as versatile as that of a human
being.  
     
Yet, as emphasised in \cite{Donald1999}, a lesson from real brain dynamics is that consciousness cannot be described statically, by assigning labels to mental states as in equation (\ref{EqualAmplitudeSplitting}), but that neural activity is required, like the switching between firing and resting states of a neuron. It makes a great difference for model building! Neurons thus come as subsystems with two dimensions at least, and fluctuations in the number of neurons appear vastly enhanced in the dimensions of Hilbert spaces involved.

The guiding idea of the present paper is that Everett's many worlds are not all  
equivalent for an observer, but that his consciousness is \emph{physically} contained 
almost exclusively in one world. The technical basis of this approch is a theorem of order 
statistics \cite{Embrechts1997,David1981}, relating to statistical ensembles with power-law 
distribution. The largest draw, 
in that case, exceeds the second-largest by a quantity $\Delta$ of the order of the ensemble 
size taken to some power. On an Everettian world tree, the ensemble size is huge, and 
concentrated near the end of the branches. Thus the dominance of the extremal 
draw is particularly pronounced, and it occurs near the end of a branch, thus singling it out.
If the draws are for numbers of neurons, the exceedance $\Delta$ exponentiates because the 
dimensions of subsystems multiply. In fact, to see whether and how this statistical mechanism 
might be relevant for state reduction was the main guide for the constructions below. 
  
The model scenario is as follows. At an equidistant sequence of times, a quantum system is 
observed, generating $B$ branches of itself, of a number of records, and of a corresponding 
number of observer's neurons. The main interest is in the statistics of dimensions; 
therefore, system states, records, and neurons are only distinguished by an index and 
are not specified any further. The law of evolution, for a step of time, is given by a unitary 
operator. A specific form of initial state must be assumed for the scenario to unfold. This 
``objective'' part of the model dynamics is constructed in section \ref{secRS}.  
   
As to an observer's consciousness, it should be memory-based \cite{Edelman1989};
``the history of a brain’s functioning is an essential part of its nature as an object on which 
a mind supervenes'' \cite{Donald1997}. In the simplification of the model, 
memories and the history of the brain's functioning are identified with the records.  
The supervening mind is represented by neuronal activity drawing on memories, i.e., records.
The implementation of this drawing, by a (pseudo-)random cascade of neuronal activity, 
serves two purposes: generating a draw from a power-law distribution, and composing in one 
draw a conscious history---by ``recalling'' what happened at a certain time, then what happened 
before and after, and before and after that again, and so on. This ``subjective'' part of the 
model dynamics is constructed in section \ref{secObserversQuest}. 

At several points in the construction of the evolution operator, random draws are made. As 
mentioned already, these are supposed to approximate some complex deterministic law of 
evolution, and they are made ``once and for all times''. 
The evolution operator is the same at all times. 

Some conclusions are given in section \ref{Conclusions}, and a technically convenient restriction of the dynamics is defined in appendix \ref{secAvoidingLoops}.

\section{Records and objective dynamics\label{secRS}}

\subsection{A paradigm: Griffiths' models of histories}

An elaborate version of the Copenhagen interpretation is the formalism of consistent histories \cite{Griffiths2002}. The notion of measurement on a quantum system is modified to that of a property (selected from an available spectrum) which the system has, irrespective of whether an observer is in place. Those properties a system has at times $t_0,t_1,\ldots, t_n$, while no property should be imagined for intermediate times. By a generalised Born's rule, probabilities are defined for sequences of the available properties (histories of the system) to occur. There are constraints, involving the time evolution operator, to be imposed on the sort of properties that can form a history. In order to demonstrate the constraints, Griffiths devises a number of toy models by immediately constructing operators of time-evolution, rather than Hamiltonians. For this purpose, the tensor-product structure of the so-called history Hilbert space turns out to be quite convenient. Moreover, models for simultaneously running processes in a time interval can be simply constructed as products of unitaries.  

Although the intention in \cite{Griffiths2002} is to keep observers out of the theory, it certainly is a reasonable approximation for an observer model as well to consider awareness only at times $t_0,t_1,\ldots, t_n$ while leaving unspecified observer's state in between. In the model to be constructed, the times will be equidistant. The operator of evolution from an instant to the next will be the product $U_\mathrm{wit\,2}U_\mathrm{con}U_\mathrm{wit\,1}U_\mathrm{orb}U_\mathrm{age}$, their roles being to increase the ages of records, move the system along its (branching) orbit, create first half of witnessing records, run awareness cascade, and create second half of witnessing records. Each of these is a product of a large but finite number of unitary factors.   

The tensor product structure of the space of states, which in Griffiths' formalism emerges from the notion of history Hilbert space, is a convenient element of model building independently of that notion, because it facilitates the construction of operators that manifestly commute.   Below, the tensor product will describe a reservoir of potential records and their neuronal counterparts, each of which being treated as a subsystem. By construction of the evolution operator, every branch will consist of its own collection of subsystems. This makes the numbers of subsystems very large, raising the question of whether a ``multiverse'' in the classical sense is tacitly assumed for the model. It should therefore be noted that the subsystems can be mathematically identified with degrees of freedom of a conventional Hilbert space. For example, any unitary space of dimension $2^n$ can be rewritten as a tensor product by expressing the index of basis vectors in binary form and identifying 
$$
   |i_1,\ldots,i_n\rangle \equiv |i_1\rangle \cdots |i_n\rangle \qquad i_k=0,1
$$  
For the counting of dimensions, both versions are equivalent. 
In order to show that one branch nearly exhausts all dynamical dimension of awareness, what matters is the quantity of records and of neuronal response. The only quality of relevance is the date at which a record is created. For the model it therefore suffices to distinguish records by an unspecified index, and to make their age the only stored content.

\subsection{Structure of state vectors; dating of records\label{secAgeing}}

The subsystems of the model are ``objective'' records and ``subjective'' bits of
mental processing. For the records, two classes are assumed. Those in class ${\cal O}$ (orbits) induce further records in the course of evolution, giving rise to branching world lines. The branches emanating from a record $k\in{\cal O}$ are collected in a set ${\cal B}(k)$ of $B$ elements. They are envisioned as predetermined and (due to complex dynamics of real macroscopic systems) pseudorandom. They are defined here\footnote{A more technical specification, simplifying evaluation, is given in appendix \ref{secAvoidingLoops}.} by random draw, once and for all time:
\begin{equation}  \label{DefB(k)}
 {\cal B}(k) = \{j(k,1),\ldots,j(k,B)\} \mbox{ where } j(k,s) = \mbox{random draw from } 
 {\cal O}\backslash\{k\}
\end{equation}  
The other class of objective records consists of mere witnesses, holding redundant information about records in ${\cal O}$:
\begin{equation}  \label{DefW(k)}
    k\in{\cal O}\mbox{ is witnessed by all records } l \in {\cal W}(k)
\end{equation} 
All ${\cal W}(k)$ are assumed to have the same macroscopic number $W$ of elements.

Observer's neurons are associated with witnessing records, not immediately with orbital
records. The model neurons are distinguished by an index and are not specified any further.
However, it could make sense to address the same anatomical neuron at different times by 
different indices. The multiple degrees of freedom would then be provided by the metabolic 
environment.  
     
Denoting by ${\cal R}_k$ and ${\cal N}_k$ the state-vector space of a record and an observer's ``neuron'', respectively, the model Hilbert space is
$$
   {\cal H} =  \bigotimes_{k\in{\cal O}}\left({\cal R}_k \otimes    
    \bigotimes_{l\in{\cal W}(k)} \left( {\cal R}_l\otimes {\cal N}_l \right)  \right)
$$
The information to be stored in a recording subsystem is whether anything is recorded at all, 
and if so, since how many steps of evolution. The basis states of a record of the orbital kind thus are
\begin{equation}  \label{DefRecordBasisOrbit}
   |\mathrm{blank}\rangle, ~ |\mathrm{age}~m\rangle \quad m\in{\bf Z}
   ~~~~ \mbox{ for each index in } {\cal O}
\end{equation}
Observer's mental processing of a record is modelled by transitions between the firing and resting state of a ``neuron''. The basis states of a witnessing record and its mental counterpart are 
\begin{equation}  \label{DefRecordBasisWitness}
   \left\{ \begin{array}{l} |\mathrm{blank}\rangle \\ 
   |\mathrm{age}~m\rangle \quad m\in{\bf Z} \end{array}\right\} \otimes
  \left\{ \begin{array}{c} |\mathrm{rest}\rangle \\  |\mathrm{fire}\rangle \end{array}\right\}
   ~~ \mbox{for each index in } \bigcup_{k\in{\cal O}}{\cal W}(k) 
\end{equation}
The ages of records could be limited to an observer's lifetime, as it was done in \cite{Polley2012}, but the infinite dimension implicit in (\ref{DefRecordBasisOrbit}) and (\ref{DefRecordBasisWitness}) is harmless with finite products of unitary operators, and avoids an unnecessary degree of subjectivity in the model. 
       
For all recording subsystems, an ageing operation is defined:
\begin{equation}  \label{DefUage}
   U_\mathrm{age} = \prod_{k\in{\cal O}} U_\mathrm{age}(k) 
                                  \prod_{i\in{\cal W}(k)}U_\mathrm{age}(i) 
\end{equation}
where for record number $r$
$$
   \begin{array}{l}
  U_\mathrm{age}(r)|\mathrm{blank}\rangle_r = |\mathrm{blank}\rangle_r \\
  U_\mathrm{age}(r)|m\rangle_r = |m+1\rangle_r 
   \end{array} 
$$

\subsection{Preferred initial state\label{secInitialState}}

A real observer's identity derives from a single DNA molecule. For a model of an observer's 
history, this is taken here as justification for considering exclusively the evolution from an 
initial state in which one orbital record $k_0$ and its witnesses ${\cal W}(k_0)$ are in the 
zero-age state while all other records are in their blank states. The choice of a zero-age 
record determines observer's entire history. 
It is the only ``seed'' for all ensuing pseudo-random processes of evolution.

\subsection{Orbital branching\label{secOrbit}}

The idea is that observer's history branches at every step of evolution, like in a quantum 
measurement. A new branch is described by an index of a new record, and is not specified any
further. Imagining a new quantity being measured at every step would seem to be consistent with
this scenario.    

Deutsch \cite{Deutsch1999} showed, for systems with sufficiently many degrees of
freedom, that Born's rule for superpositions with coefficients more general than in equation
(\ref{EqualAmplitudeSplitting}) can be reduced to the
equal-amplitude case, providing the unitarity of any physical transformation is taken for
granted. In the model to be constructed, evolution will be unitary. Therefore,
invoking Deutsch's argument, only branching into equal-amplitude superpositions will be
considered.

Under the condition that record $k$ is older than zero, and that all records to 
which the orbit possibly continues are blank, the orbit does continue as a superposition of 
zero-age states of the records of address set ${\cal B}(k)$. Else, the identity operation is carried out. The corresponding evolution operator, specific to point $k$ on an orbit, is defined using the following basis of the \emph{partial tensor product} relating to the records of the 
set ${\cal B}(k)$.  
\begin{equation}  \label{PartialBasis}
    \begin{array}{l}
    |\Psi_0\rangle = \prod_{l\in{\cal B}(k)} |\mathrm{~blank}\rangle_l \\[3mm]
    |\Psi_l\rangle = \Big(|0\rangle\langle\mathrm{blank}|\Big)_l|\Psi_0\rangle 
    \qquad l\in{\cal B}(k)
    \end{array}
\end{equation}
That is, one basis vector has all records of ${\cal B}(k)$ in the blank 
state, while the remaining have one record promoted to the zero-age state.
The subspace orthogonal to $|\Psi_0\rangle$, \ldots, $|\Psi_B\rangle$
is spanned by product vectors with more than one record in a zero-age state or with records
in higher-age states. 

The idea of equal-amplitude branching from point $k$ is that $|\Psi_0\rangle$ should evolve 
into a superpositon of $|\Psi_1\rangle$ to $|\Psi_B\rangle$; in the basis (\ref{PartialBasis}),  
\begin{equation}   \label{SplittingShorthand}
     \left(\begin{array}{c} 1 \\ 0 \\ \vdots \\ 0 \end{array} \right) \longrightarrow
     \frac1{\sqrt B} \left(\begin{array}{c} 0 \\ 1 \\ \vdots \\ 1 \end{array} \right)
\end{equation} 
A convenient way of completing this to define a unitary operator is to use a Fourier basis in 
$B$ dimensions, 
$$
  F_m = \frac1{\sqrt B}\left(\begin{array}{c} \alpha_m^0 \\ \alpha_m^1 \\ \vdots \\ 
    \alpha_m^{B-1} \end{array} \right) \qquad \alpha_m = \exp\frac{2\pi i m}{B} 
    \qquad m = 0,\ldots B-1  
$$   
Relating to the basis (\ref{PartialBasis}), and using the $F_m$ as $B$-dimensional column vectors, a branching operation can be defined using the $(B+1)\times(B+1)$ matrix 
$$
    S = \left(\begin{array}{ccccc} 0 & 1 & 0 & \cdots & 0 \\
             F_0 & 0 & F_1 & \cdots & F_{B-1} \end{array} \right) 
$$ 
whose columns form an orthonormal set. 
Conditioning on $\mathrm{age} = 1$ of record $k$, the factor of orbital evolution triggered by
this record is then given by
\begin{equation}  \label{DefUorb(k)} 
   U_\mathrm{orb}(k) = 1 + \left( \sum_{n,n'=0}^B |\Psi_n\rangle 
  (S_{nn'}-\delta_{nn'}) \langle\Psi_{n'}|\right)_{{\cal B}(k)}  
  |\mathrm{age}~1\rangle_k \langle \mathrm{age}~1|_k
\end{equation}
The bracket reduces to zero, in particular, when branching from $k$ has occurred previously in the evolution, so that, by subsequent ageing of non-blank records, 
any zero-age components of records have been promoted to higher-age components; 
cf.\ (\ref{PartialBasis}). In this way, repeated branching from the same point as well as loops of evolution are avoided. Technically, however, an additional means of avoiding loops (appendix \ref{secAvoidingLoops}) facilitates the evaluation of evolution. 
The global operator of orbital evolution is 
\begin{equation}  \label{DefUorb}
   U_\mathrm{orb} = \prod_{k\in{\cal O}} U_\mathrm{orb}(k)
\end{equation}

\subsection{Witnessing}

The basis states of the witnessing records are defined in (\ref{DefRecordBasisWitness}).
In order to represent (\ref{DefW(k)}) by an evolution operator, orbital records of zero age
are assumed to induce a change of witnessing records from blank to zero-age. The relevant part of the operation is, in self-explaining notation, 
\begin{equation}  \label{UwitSimpl}
  \left( \prod_{l\in{\cal W}(k)} \Big(|0\rangle \langle\mathrm{blank}|\Big)_l\right) 
  \Big( |0\rangle \langle 0|\Big)_k 
\end{equation}     
For the sake of unitarity, however, this needs to be complemented by further operations, 
although these will never become effective in the evolution of an initial state as defined in section \ref{secInitialState} and as age-promoted by the operators of section \ref{secAgeing}. 
A complemented version of the operator above would be 
$$
 1 + \left( -1 + \prod_{l\in{\cal W}(k)} \left[|0\rangle \langle\mathrm{blank}|
  + |\mathrm{blank}\rangle \langle 0| + \sum_{m\neq 0}|m\rangle\langle m|\right]_l 
   \right) \Big( |0\rangle \langle 0|\Big)_k
$$
However, witnessing records just created can be read and processed by an observer within the 
same step of evolution. It will be essential for the functioning of a stochastic mechanism below (section \ref{secObserversQuest}) that half of the witnessing records are created before the observer might immediately address them, while the other half is created thereafter. For this purpose, let the addresses of witnessing records be split into subsets of equal size,
$$
    {\cal W}(k) = {\cal W}_1(k) \cup {\cal W}_2(k) ~~~~~~  
    {\cal W}_1(k) \cap {\cal W}_2(k) = \emptyset
$$   
The two corresponding witness-generating evolution operators are
\begin{equation}  \label{DefUwit}
   U_\mathrm{wit\,1} = \prod_{k\in{\cal O}} U_\mathrm{wit\,1}(k) ~~~~~~~~~~~~~~~~~~
   U_\mathrm{wit\,2} = \prod_{k\in{\cal O}} U_\mathrm{wit\,2}(k)
\end{equation}
where \vspace*{-4mm}  
$$
 ~~~~~~~~  \begin{array}{l}
 U_\mathrm{wit\,1,2}(k) ~ = ~ 1 ~ +  \\[1mm]  \displaystyle
  \left( -1 + \!\!\! \prod_{l\in{\cal W}_{1,2}(k)} \left[|0\rangle \langle\mathrm{blank}|
  + |\mathrm{blank}\rangle \langle 0| + \sum_{m \neq 0}|m\rangle\langle m|\right]_l 
   \right) \Big( |0\rangle \langle 0|\Big)_k
 \end{array}
$$

\section{Observer's mental programme\label{secObserversQuest}}

Consider an observer who is constantly trying to assemble his records into a coherent temporal 
sequence. One way of doing this would be to see what happened at the middle $a/2$ of his life 
at age $a$, by seeking an appropriate record; then what happend a quarter before and after, at 
$a/4$ and $3a/4$, then at multiples of $a/8$, and so forth. 
This defines a cascade of increasing temporal resolution. For 
``coherence'', connection by a logical AND is required. It would fit in with the spirit (not 
with the technical detail) of Tononi's concept of consciousness as Integrated Information 
\cite{Tononi2008}: 
``Phenomenologically, every experience is an integrated whole, one that means what it means by 
virtue of being one, \ldots''. Moreover, memory 
becomes a fundamental constituent of consciousness \cite{Edelman1989} in this way.

\subsection{Generating the power-law statistics\label{secGenPowStat}}

The idea for generating power-law statistics, within one step of evolution, is as follows.   
While the AND condition is satisfied, the cascade of records addressed grows like $2^l$ where 
$l$ is the generation number. By the scanning procedure to be constructed, records of non-zero 
ages will be found with probability 1 whenever addressed, but records of zero age (being 
created within the same step and representing the ``present'') will only be found with 
probability 1/2. The entire cascade is stopped when the quest for a coherent picture of past 
and present fails, for which the probability is 1/2 in every generation. This is a standard 
mechanism for generating power law statistics \cite{SimkinRoychowdhury2006}, here with exponent 
$-1$ for the cumulative distribution function since a number greater than $n=2^{l+1}-1$ (sum of 
generations) is obtained with probability $\frac12(n+1)^{-1}$.

\subsection{Recursive construction of awareness cascade} 

The unitary operator to be constructed in this section will go through all possible cascades 
of records of ages $a/2$, multiples of $a/4$, of $a/8$ and so on, looking for a randomly 
chosen witnessing record $i$ in every ${\cal W}(k)$ of the cascade. It should be noted again 
that all random draws are made once and for all times. Parameter $a$ will be an eigenvalue of 
observer's age operator, defined in section \ref{secObserversAge}.  

We begin by constructing the $l$th generation of the cascade. Consider a set $g$ of addresses 
given by pairs $(k,i)$ with
\begin{equation}  \label{ikjDef}
  \left. \begin{array}{rcl} 
      k(j) & = & \mbox{label of an orbital record} \\
      & & \mbox{restricted by } k(j) < k(j') \mbox{ for } j < j' 
      \\[3mm]
      i(j) & = & \mbox{random draw from ${\cal W}(k(j))$}
  \end{array} \right\} \mbox{ for }j = 1,\ldots,2^l
\end{equation}
The ordering of the $k$ is for technical convenience; permutations are taken into account when  
assigning ages, as below. Denote the collection of all possible sets of the above form by 
$$
   {\cal G}(l) = \big\{ \mbox{all possible $g$ of the form (\ref{ikjDef})} \big\}
$$
The random draws are understood to be \emph{independent for different} $g$. 
The elementary projection on which the scanning operation is based is 
\begin{equation}  \label{PaikDefinition}
  \Big( |m\rangle \langle m|\Big)_i = \mbox{projection on age $m$ of record $i$}
\end{equation}    
Below, fractional ages are converted to integers by the ceiling function $\lceil~\rceil$. 
To enable the scanning of all combinations of ages and records, let us define
\begin{equation}  \label{alphaDef}
    \alpha = \mbox{sequence consisting of ages }\lceil 2^{-l}(j-1)a \rceil,  
                j = 1,\ldots,2^l, \mbox{ reordered}
\end{equation}
This distinguishes permutations of different ages, but not of equal ages. 
Denote the collection of all sequences of the form (\ref{alphaDef}) by
$$
   {\cal A}(l) = \big\{ \mbox{all possible $\alpha$ of the form (\ref{alphaDef})} \big\}
$$
The projection operator testing whether the records given by $g$ have ages as given by 
$\alpha$ is 
\begin{equation}  \label{DefP(g,a)}
   P(g,\alpha) = 
  \left[\prod_{j=1}^{2^l} \Big( |\alpha(j)\rangle \langle \alpha(j)|\Big)_{i(j)}\right]  
  \left[\prod_{i \notin g} \Big( |\mbox{blank}\rangle \langle \mbox{blank}|\Big)_i\right] 
\end{equation}   
where $i(j)$ in the first bracket denotes elements of $g$. These projectors are mutually
orthogonal,
\begin{equation}  \label{OrthogonalP(g,a)}
    P(g,\alpha) P(g',\alpha') = 0 ~~ \mbox{ if } g\neq g' \mbox{ or } \alpha\neq \alpha' 
\end{equation} 
To show this, consider $g \neq g'$. Let $i$ be an index in $g$ but not in $g'$. Then in 
$P(g,\alpha)$ we have a projector $( |m\rangle \langle m|)_i$ while in $P(g',\alpha')$ we have 
$( |\mbox{blank}\rangle \langle \mbox{blank}|)_i$ instead. The product of these two is zero 
already. Secondly, consider the case of $g=g'$ and $\alpha\neq\alpha'$. 
Let $j_0$ be an index for which $\alpha(j_0)\neq \alpha'(j_0)$. Now the projectors are
orthogonal because they project on different ages for record $i(j_0)$.

The idea for the modelling of observer's neuronal reaction is as follows. 
In the subspace where the test by $P(g,\alpha)$ is positive (all $p$ give 1) the observer 
notices it by some neural activity, and the next generation of the scanning process takes 
place. In the subspace where the test is negative (some $p$ give 0) nothing happens; evolution
reduces to an identity operation. The neural activity is modelled by 
2-dimensional rotations $\sigma_i$ in the counterparts ${\cal N}_i$ of the records. 
In the subspace where $g$ tests positive, the collective rotations are, with obvious 
assignment to the tensorial factors,  
$$
    \sigma(g) = \prod_{i \in g} \sigma_i
$$
The awareness cascade, running within a step of evolution, is conveniently constructed 
recursively, downward from higher to lower resolutions of time. This is enabled by the fact 
that after many generations the finite contents of the address sets ${\cal W}(k)$ will be 
exhausted. So there is a maximum $L$ for the generation number $l$, determined by the other 
parameters of the model. The counting of the generations will be upward here as usual, 
beginning with $l=1$ at age $a/2$. The recursion is initialised by 
\begin{equation}  \label{UawaInitial}
      U_\mathrm{awa}(L+1,a) = 1 
\end{equation}
and proceeds by
\begin{equation}  \label{UawaRecursion}
    U_\mathrm{awa}(l,a) = 1 +  \sum_{g\in{\cal G}(l)} \sum_{\alpha\in{\cal A}(l)} 
                 \Big( -1 + U_\mathrm{awa}(l+1,a) \sigma(g) \Big)  P(g,\alpha) 
\end{equation}
The awareness operator for the completed cascade is $ U_\mathrm{awa}(1,a)$. Defining it by 
recursion is only a convenient way of representing the algebraic structure. In application
to a state vector, projections of the various generations automatically occur in the natural 
order, $l=1,\ldots,L$.   

In order to show that (\ref{UawaRecursion}) indeed defines a unitary operator, first note
that $P(g,\alpha)$ only consists of projections on the ages of witnessing records, so that 
basis states of the form (\ref{DefRecordBasisWitness}) are eigenstates of the projectors. 
All age projectors commute among themselves, and commute with the $\sigma$ 
operations because these do not act on records. It follows, starting from 
(\ref{UawaInitial}) and going through (\ref{UawaRecursion}), that the projectors commute
with the $U_\mathrm{awa}$ of all generations. Using (\ref{OrthogonalP(g,a)}), unitarity in the 
form $U_\mathrm{awa}(l,a)^\dag U_\mathrm{awa}(l,a) = 1$ can then be shown by 
straightforward algebra.

\subsection{Observer's age and conscious history\label{secObserversAge}}
 
Parameter $a$ of the preceding section is identified here as an eigenvalue of observer's 
age operator. It suffices to assign an age to any tensor product of basis vectors as defined in 
(\ref{DefRecordBasisWitness}). Assigning 0 to the ``blank'' state here, any basis state of 
record $i$ has an age value $a_i$. The age operator is defined by  
$$
   A \prod_i |a_i\rangle = ( \max a_i) \prod_i |a_i\rangle 
$$   
Observer's lifetime can be taken into account by restricting neuronal activity to ages 
$ a \leq T$, by including an operator factor $\Theta(T-A)$. 
Thus, the final expression for the evolution operator of observer's consciousness is
\begin{equation} \label{DefUconsc}
   U_\mathrm{con} = 1 + \Big(-1 + U_\mathrm{awa}(1,A)\Big)\Theta(T-A)
\end{equation}   
The complete evolution operator is a product of the factors constructed above.
In a new step of evolution, ages of all records are increased by one unit. 
Next, orbital records develop. The creation of witnessing records and their processing by the 
observer (operations that do not commute) are assumed to be intertwined in such a way that 
unitarity is manifestly preserved. The full evolution operator of the model is
\begin{equation}  \label{DefUmodel}
 U = U_\mathrm{wit\,2} ~ U_\mathrm{con} ~  U_\mathrm{wit\,1} ~ U_\mathrm{orb} ~ 
 U_\mathrm{age}            
\end{equation}

\subsection{Verifying the Scenario\label{ScenarioRecovered}}

\subsubsection{Structure of branches}

We start out from a product state as specified in section \ref{secInitialState} and repeatedly apply the evolution operator (\ref{DefUmodel}). Clearly, since product states form a basis, we can always write the resulting states as superpositions of products; however, we wish to show that only a superposition of special products emerges, which will be regarded as the ``branches'' of the wave function. After $a$ applications of the evolution operator $U$, the properties of those product states are as follows. 
\begin{enumerate}
\item In each branch there is precisely one orbital record of age $0$. 
\item For each of the ages $0,\ldots,a$, there is one set ${\cal W}(k)$ of 
      witnessing records in the corresponding eigenstates of age; all other witnesses 
      are blank. 
\item Neuronal states and record states factorise (do not entangle). 
\end{enumerate}  
The initial state has these properties with $a=0$ by definition. Let us assume then that $a$ 
applications of $U$ have produced a superposition of product states with properties 1-3. When 
$U$ is applied once more, it suffices by linearity to consider the action on any of the product 
states. The first action is to increase by 1 the ages of all non-blank records. 
There is now for each of the ages $1,\ldots,a+1$ precisely one set ${\cal W}(k)$ of 
witnessing records in the corresponding eigenstates of age, while all other witnesses 
are blank; witnesses of age $0$ are missing so far.    
Also, a single orbital record of age 1 is generated from the previous one of age 0; let its  
address be $k_1$. Now applying $U_\mathrm{orb}$, as defined in (\ref{DefUorb}), only the factor
with $k=k_1$ can have an effect since the projection on age 1 gives zero for all other $k$.  
Nontrivial action of $U_\mathrm{orb}(k_1)$ requires all records in ${\cal B}(k_1)$ to be blank, 
which would not be true if the system had been on any of those points before. Invoking the loop-
avoiding specification of ${\cal B}(k_1)$, as given in appendix \ref{secAvoidingLoops}, we can 
regard this condition as satisfied within observer's lifetime. The action of 
$U_\mathrm{orb}(k_1)$ then consists in creating a new superposition of products, with a single 
zero-age orbital record in each of them, as expressed in (\ref{SplittingShorthand}). Property 1 
holds in each of these products. For the remaining operations of $U$ it suffices by linearity 
again to apply them only on the product states just created by $U_\mathrm{orb}(k_1)$.  
Let $k_0$ be the single zero-age orbital record in one of them. Then, of all 
witness-generating factors of (\ref{DefUwit}), only $U_\mathrm{wit\,1}(k_0)$ and 
$U_\mathrm{wit\,2}(k_0)$ act nontrivially, due to the conditioning on zero-age. 
The records of the sets ${\cal W}_1(k_0)$ and ${\cal W}_2(k_0)$ are in blank states before this 
action, because orbital point $k_0$ was not visited before, so the simplified expression 
(\ref{UwitSimpl}) applies, and the records of ${\cal W}_1(k_0)$ are transformed from blank to 
zero-age states. Thus, $U_\mathrm{wit\,1}(k_0)$ generates the first half of zero-age witnessing 
records that were missing so far from the full range of ages.
Next comes the action of the action of $U_\mathrm{con}$, defined in (\ref{DefUconsc}). It is 
the only factor of evolution which could affect property 3. It  
consists in making certain neuronal factors rotate if certain projections on the ages of 
records are nonzero, and no action else. All records are in definite ages or 
blank, so the projections of $U_\mathrm{con}$ preserve the product form of the state vector.
The neuronal rotations preserve the product form by construction. Hence, property 3 continues 
to hold. Finally, the second half of zero-age witnessing records is generated by 
$U_\mathrm{wit\,2}(k_0)$, so property 2 holds as well after $a+1$ applications of the 
evolution operator.  

\subsubsection{Awareness cascades}

To evaluate the awareness cascades encoded in $U_\mathrm{con}$, defined in 
(\ref{DefUconsc}), assume that observer's age is $a<T$ so that $U_\mathrm{awa}(1,a)$ 
applies. The projection operators $P(g,\alpha)$ of (\ref{UawaRecursion}), using 
(\ref{ikjDef}) and (\ref{alphaDef}) for $l=1$, test for randomly chosen records
in ${\cal W}(k(0))$ and ${\cal W}(k(1))$, with ages $0$ and $\lceil a/2\rceil$ or the 
permutation of these, while the pair of $k(0)$ and $k(1)$ is ordered. By property 2 above,
the product state (or branch) being considered has records of one set ${\cal W}(k_0)$ at age 
$0$ and of one set ${\cal W}(k_1)$ at age $\lceil a/2\rceil$. Hence,
the only successful projection $P(g,\alpha)$ can be for $g=(k_0,k_1)$ and 
$\alpha=(0,\lceil a/2\rceil)$ or for $g=(k_1,k_0)$ and $\alpha=(\lceil a/2\rceil,0)$, 
depending on which of the addresses $k_0$ or $k_1$ is smaller. Only one term, at most, 
contributes to the sum over $g$ and $\alpha$ in (\ref{UawaRecursion}).     
The test for $k_1$ with age $\lceil a/2\rceil$ will be positive, since all
witnesses of ages $1$ to $a$ have been created during the preceding steps of evolution. 
However, witnesses of age $0$ are created half before the action of $U_\mathrm{con}$ and half 
thereafter. If the randomly chosen record from ${\cal W}(k_0)$ is contained in the first half, 
${\cal W}_1(k_0)$, it is created by $U_\mathrm{wit\,1}$ before the action of $U_\mathrm{con}$, 
so the test will result in $P=1$ in equation (\ref{UawaRecursion}); if it is created by 
$U_\mathrm{wit\,2}$ instead, the test will result in $P=0$. In the latter case, 
$U_\mathrm{awa}(1,a)$ reduces to the identity operation. 
In the case of $P=1$, the neurons associated with the witnesses for $k_0$ or $k_1$
become active through the $\sigma$ factor, and the second generation of the cascade comes into
action through $U_\mathrm{awa}(2,a)$. 
The argument repeats: As a consequence of property 2, at most one combination $(g,\alpha)$ 
contributes to the sum (\ref{UawaRecursion}) for $U_\mathrm{awa}(2,a)$, namely that combination 
in which the records collected in $g$ are tested for the ages they actually have on the branch 
considered. The zero-age orbital record $k_0$ and its witnesses are the same for all 
generations of the cascade, but for each $l$ a new randomly chosen witness is tested.         
Projection $P(g,\alpha)$ reduces to $1$ if that witness happens to be created by 
$U_\mathrm{wit\,1}$, while it reduces to $0$ if it is created by $U_\mathrm{wit\,2}$. The 
probability for the cascade to continue is $1/2$ in every generation. Witnesses of higher age 
always test positive, as they have been created in the preceding steps of evolution.

\subsubsection{Statistics of dimensions of conscious subspaces\label{DimensionStatistics}}

If the observer lives to age $T$, the number of orbital points on his world-tree is 
$$
   N = \frac{B^{T+1}-1}{B-1}
$$
This is also the number of statistically independent awareness cascades, as we now show.
By definition (\ref{ikjDef}), a new series of random draws is made for every 
sequence $g$, a selection of orbital addresses. This definition does not a priori relate to a 
specific time, but its occurrence in the projector $P(g,\alpha)$ of the evolution operator, 
defined in (\ref{DefP(g,a)}), combines it with a sequence of observer's ages. 
We intend to show the following: If projections with the same sequence, $g_1=g_2$, give 
positive results for two points on observer's world-tree, those points must be equal.

Let $a_1$ and $a_2$ be observer's ages at the two points, and let $k_1$ and $k_2$ be the 
orbital points of zero age, the ``present'' points. By (\ref{DefP(g,a)}) and (\ref{alphaDef}), 
the present point is always contained in $g$, so $k_1\in g_1$ in particular. This here implies 
$k_1\in g_2$. Since $P(g_2,\alpha_2)$ is assumed to test positive, $k_1$ must be an orbital 
record on the branch leading to $k_2$, so it either coincides with $k_2$ or is a record of age 
greater than zero. In the latter case, it must have been the zero-age record at an earlier 
age of the observer. Hence, $a_1<a_2$, unless $k_1=k_2$. Exchanging 1 and 2 in the 
argument, we find $a_1>a_2$ unless $k_1=k_2$. This implies $k_1=k_2$ and $a_1=a_2$. 
   
For the number of neurons activated in a cascade (section \ref{secGenPowStat}) the 
probability distribution is a power law characterised by exponent $-1$. Hence, by a theorem of 
order statistics \cite{Embrechts1997}, the largest number of neurons activated 
exceeds the second-largest by a quantity of order $N$. More precisely, using notation of 
\cite{Embrechts1997} corollary 4.2.13, given an ensemble of size $N$ of random draws with the 
power-law distribution, we have for the difference between the largest draw $X_{1,N}$ and the 
second-largest $X_{2,N}$  
\begin{equation}   \label{FrechetSeparation}
    X_{1,N}-X_{2,N} = N \, Y \qquad \mbox{$Y$ = random variable independent of $N$} 
\end{equation}
The distribution of $Y$ is non-singular. 
The dimension of the active neuronal subspace in the extremal branch is $2^{X_{1,N}}$. 
The total dimension of active neuronal subspaces in all other branches is bounded by
$2^{X_{2,N}}N$. For the latter to be larger than the former, the probability is
$$
    P(2^{NY}<N) = P(Y<N^{-1}\log_2N) = \mbox{negligible}
$$  
By a comfortable margin, an observer can expect to find his world-line well-defined, providing 
it is indeed the \emph{dimension} of awareness that matters, rather than the number of neurons.

Two arguments in favour of the dimension are at hand. The simplest is Fermi's Golden Rule, 
although it relies on state reduction and so goes beyond the framework of the model; any 
transition probabilities into observer's subspace of awareness would be proportional to the 
dimension of the subspace.      
The other argument uses a change of basis in the union of conscious subspaces of \emph{all} 
branches. Let $|1\rangle,\ldots,|N\rangle$ be a basis for the extremal branch, and 
$|N+1\rangle,\ldots,|N+M\rangle$ a basis for the remaining branches. We know that $M$ is tiny 
in comparison to $N$. Now consider instead a Fourier basis, which consists of superpositions
of all $|1\rangle,\ldots,|N+M\rangle$ with equal amplitudes but different phases. In each of 
the new basis vectors, properties of the non-extremal branches only occur in a tiny
component. The situation is now similar to that of an electron bound to a proton on earth. 
It resides by $10^{-10^{18}}$ of its wavefunction behind the moon, but we still regard it as
an electron on earth.

\subsection{Analysing states in a time-local basis\label{secLocalBasis}}

In order to analyse the properties of a state vector, it must be represented in a particular basis, such as the eigenbasis of an observable. For some applications, like representing dynamics in the Heisenberg picture, the basis may conveniently be chosen time-dependent. 

As to the state vectors of observer's neurons, we have so far used a global basis which applies to all branches, and in which the evolution operator is constant. In this way, observer's entire experience emerges in a single step of evolution near the end of his lifetime. Observer's mental reactions thus appear highly non-local. 
However, when analysed in a time-dependent basis adapted to the evolution in one particular branch, observer's reactions occur simultaneously with the creation of the records, while the operator of evolution appears to change in a random way after each observation. This is trivial mathematically, but not physically. 
Consider a section of evolution of a neuronal subsystem, 
$$
   \left(\begin{array}{c} x_1 \\ x_2 \end{array}\right) \stackrel{1}{\longrightarrow}
   \left(\begin{array}{c} x_1 \\ x_2 \end{array}\right) \stackrel{\sigma}{\longrightarrow}
   \left(\begin{array}{c} y_1 \\ y_2 \end{array}\right) 
$$
where entries relate to some initially chosen basis, and where $\sigma$ is a unitary $2\times 2$ matrix. Only in the second step something appears to happen here. Changing the basis for the second state vector such that
$$
    \left(\begin{array}{c} x_1 \\ x_2 \end{array}\right)_\mathrm{old~basis}
 =  \sigma^{-1}\left(\begin{array}{c} y_1 \\ y_2 \end{array}\right)_\mathrm{new~basis}
$$
the section of evolution takes the form
$$
   \left(\begin{array}{c} x_1 \\ x_2 \end{array}\right) \stackrel{\sigma}{\longrightarrow}
   \left(\begin{array}{c} y_1 \\ y_2 \end{array}\right) \stackrel{1}{\longrightarrow}
   \left(\begin{array}{c} y_1 \\ y_2 \end{array}\right) 
$$
The step of evolution where change appears can be shifted to any position in the 
sequence. Due to the tensor-product structure of the model's branches, the argument applies 
separately to all neurons involved. It is thus \emph{possible} to choose a basis in which 
observer's neuronal reactions appear local, but it is a different choice in each branch, and 
the reason for the choice cannot be found in the state of records at the given time. In this 
sense, the choice appears to be intrinsically random.

\section{Conclusions\label{Conclusions}}

It has been demonstrated for a toy model of a quantum system with conscious observer, 
that a unitary evolution operator, repeatedly applied to an appropriate initial state, 
can accomplish two things: gather information about an observer's world-tree,
and perform a random draw on the world-tree so as to single out a world-line of extreme
awareness. A theorem, known from order statistics, about the dominance of the extreme in a 
power-law ensemble plays a central role. 

What the model \emph{avoids} to do is giving up fundamental linearity, and
introducing fundamental stochastics. The framework is state vectors and unitary evolution
under a constant law. Yet certain vectors evolve pseudo-stochastically.        

The role of time and causality in the model is precarious, inevitably so under
the working hypothesis that a world-line should be determined by an extremal
draw on a world-tree.      
As was shown in section \ref{secLocalBasis}, the model reproduces the usual
scenario of alternating Schr\"odinger-type and Born-type evolution when
represented in an appropriate basis. Observer's mental reactions then appear at the same
instant as the generation of records, but the choice of the basis appears indeterminate at
that instant. The model resolves that indeterminacy by omitting any erosion of
``witnessing records'', keeping them readable throughout observer's lifetime. 
Can this be true for more realistic ``witnessing records'', or is this the point where
attempts at realistic modifications of the model must fail? 

Non-universality of time could be an argument in favour of the optimistic alternative. 
From Special Relativity Theory,
time is known to be observer-dependent, but only with negligible effects if 
observers move at low speed. On this basis, time is treated as universal in nonrelativistic
quantum mechanics (likewise in the preceding sections of this paper). 
But quantum mechanics provides its own path to special relativity, in the sense that it 
enables pre-relativistic derivations of the Dirac equation \cite{Zee1991,Polley2001}; 
it might also provide its own version of observer-dependent time. The existence of two 
modes of evolution, Schr\"odinger and Born, might be an indication of it. 
  
Having recovered the stochastic appearance of measurements in section \ref{secLocalBasis}
by referring to a specific basis, we may have some freedom in reinterpreting the 
evolution \emph{operator} as something more general. Since logics is always part of a natural 
law, and conceptually more general, could the role of the operator be to generate a logical 
structure of which time evolution is only a representative in a particular basis? 
In elementary cases like those described by a Dirac equation, the law of motion is close to 
mere logics of nearest neighbours \cite{Polley2001}, so ``space-time'' might indeed reduce 
to ``space-logics''. For quantum systems with great complexity, like an 
elementary system coupled to a conscious observer, logical implications might depend on many 
conditions, and could be halted as long as some conditions were not met by the state vector. 

In different ways, ``halted'' evolutions are also considered in other scenarios
of quantum measurement. Stapp \cite{Stapp1993} proposed an interaction between mind and matter 
based on the quantum Zeno effect; it would keep observer's attention focussed on one outcome in
a measurement, but a side effect would be the halting of processes in matter under observation.
With consistent histories \cite{Griffiths2002}, there is a copy of Hilbert space assigned to 
each of the times $t_1,\ldots,t_n$ of measurement. In the present model, an analogue of such 
``history Hilbert spaces'' may be seen in the subspaces defined by the written states of 
records at a time $t_k$.   
  
The modelling of 
consciousness by a coherent, logically conjunctive neuronal activity appears to be in 
the \emph{spirit} of Integrated Information \cite{Tononi2008}. As a \emph{measure} of 
consciousness, however, section \ref{DimensionStatistics} of the present paper
suggests to take the total dimension of the subspace of neuronal activity, which is very
different from the entropy-based measure suggested in \cite{Tononi2008}. By taking logarithms 
of dimensions, an elementary relation like that of one subspace covering 
the union of many other subspaces becomes almost invisible.

If the model scenario could indeed be extended to more realistic systems and
observers, it would suggest an easier intuitive look on state vectors, and on the persistent
problem of ``the Now'' \cite{Mermin2013}. Intuitively, states of superposition have always been
associated with potentialities for a quantum system, but the need for an actuality
seemed to make it an insufficient characterisation. The model scenario suggests to identify
actuality with that potentiality which involves an extremal degree of awareness. 
It is generated in one step of logical evolution, so an observer's impression of his entire 
experience as one shifting moment would seem less surprising.

\begin{appendix}

\section{Suspended orbital return\label{secAvoidingLoops}}

Presumably, the probability for an orbital point to be visited twice under the dynamics of 
section \ref{secOrbit} is negligible, but in order to enable exact statements on evolution, 
any returns of the system should be rigorously excluded for the time span of interest.
      
Loops of evolution on a branching orbit of points in ${\cal O}$ cannot be avoided entirely if 
${\cal O}$ is a finite set, but they can be avoided within an observer's lifetime. To keep 
branches apart for $T$ splittings, assume ${\cal O}$ to be decomposable into $B^T$ subsets of 
the form ${\cal J}[s]$, mutually disjoint and big enough to serve as an ensemble for a random 
draw, with $s$ a register of the form
\begin{equation}  \label{Register}
   s = [s_1,s_2,\ldots,s_T] ~~ \mbox{ where }~~ s_j \in \{1,2,\ldots,B \}
\end{equation}
Then, starting from $k\in{\cal J}[s_1,s_2,\ldots,s_T]$, define the jumping addresses 
$j(k,s)$ for branches $s = 1,\ldots,B$ by
\begin{equation}  \label{JumpingRegister}
   j(k,s) = \mbox{random draw from }[(s_2\,\mathrm{mod}\,B) + 1,s_3,\ldots,s_T,s]
\end{equation}
The cyclic permutation in the first entry serves to avoid jumping within one subset. 
Entry $s$ will remain in the register for $T$ subsequent splittings. Thereafter, 
the corresponding information is lost, allowing for inevitable loops to close. 
The addresses generated in (\ref{JumpingRegister}) are a loop-avoiding specification of the 
elements of ${\cal B}(k)$, previously defined in (\ref{DefB(k)}).  

\end{appendix}

\end{document}